\documentclass[preprint]{aastex}

\begin{document}

\title {KIC 11401845: An Eclipsing Binary with Multiperiodic Pulsations and Light Travel Time}
\author{Jae Woo Lee$^{1,2}$, Kyeongsoo Hong$^1$, Seung-Lee Kim$^{1,2}$, and Jae-Rim Koo$^1$ }
\affil{$^1$Korea Astronomy and Space Science Institute, Daejeon 34055, Korea}
\affil{$^2$Astronomy and Space Science Major, Korea University of Science and Technology, Daejeon 34113, Korea}
\email{jwlee@kasi.re.kr, kshong@kasi.re.kr, slkim@kasi.re.kr, koojr@kasi.re.kr}

\begin{abstract}
We report the ${\it Kepler}$ photometry of KIC 11401845 displaying multiperiodic pulsations, superimposed on 
binary effects. Light-curve synthesis represents that the binary star is a short-period detached system with a very low 
mass ratio of $q$ = 0.070 and filling factors of $F_1$ = 45 \% and $F_2$ = 99 \%. Multiple frequency analyses were applied 
to the light residuals after subtracting the synthetic eclipsing curve from the observed data. We detected 23 frequencies 
with signal to noise amplitude ratios larger than 4.0, of which the orbital harmonics ($f_4$, $f_6$, $f_9$, $f_{15}$) in 
the low frequency domain may originate from tidally excited modes. For the high frequencies of 13.7$-$23.8 day$^{-1}$, 
the period ratios and pulsation constants are in the ranges of $P_{\rm pul}/P_{\rm orb}$ = 0.020$-$0.034 and 
$Q$ = 0.018$-$0.031 d, respectively. These values and the position on the Hertzsprung-Russell diagram demonstrate that 
the primary component is a $\delta$ Sct pulsating star. We examined the eclipse timing variation of KIC 11401845 from 
the pulsation-subtracted data and found a delay of 56$\pm$17 s in the arrival times of the secondary eclipses relative 
to the primary eclipses. A possible explanation of the time shift may be some combination of a light-travel-time delay 
of about 34 s and a very small eccentricity of $e \cos {\omega} <$ 0.0002. This result represents the first measurement of 
the R{\o}mer delay in non-compact binaries. 
\end{abstract}

{}

\section{INTRODUCTION}

The {\it Kepler} satellite provides the highly precise and nearly-continuous photometric data for $\sim$ 200,000 objects that 
has helped to revolutionize the study of stars themselves, as well as extrasolar planets (Borucki et al. 2010; Koch et al. 2010). 
There are 2878 eclipsing and ellipsoidal binaries in the {\it Kepler} main field of view (Kirk et al. 2016), corresponding 
to about 1.3 \% of all observed targets. Eclipsing binaries (EBs) serve as critical tools that provide an accurate and 
direct determination of fundamental stellar parameters such as the mass and radius for each component. These data allow us 
to test stellar evolution models and to determine distances of binary systems (Torres et al. 2010). Furthermore, it is 
possible to measure precisely mid-eclipse times from binary light curves. The timing measurements are used to investigate 
a variety of physical phenomena causing the orbital period changes of EBs (Hilditch 2001; Kreiner et al. 2001). 
Such examples are mass transfer, angular momentum loss, apsidal motion in an elliptical orbit, third-body effect, and 
magnetic activity cycle. 

EBs with pulsating components are very promising objects for the study of stellar structure and evolution, because binarity 
provides useful information about the components and asteroseismology assists in probing the interiors of stars. Most of 
them have been found to be $\delta$ Sct-type pulsators of classical semi-detached Algols (Mkrtichian et al. 2004; 
Liakos \& Niarchos 2016b). The $\delta$ Sct stars are dwarfs and subgiants with spectral types A and F located in the lower 
portion of the Cepheid instability strip. They pulsate in low-order pressure ($p$) modes with typical periods of 0.02$-$0.25 days 
and amplitudes of less than 0.1 mag (Breger 2000; Rodr\'iguez \& Breger 2001). The pulsations are driven by the $\kappa$ mechanism 
mostly due to partial ionization of He II. The $\delta$ Sct variables in binaries have pulsation features similar to single 
$\delta$ Sct stars, but their pulsations can be affected by mass transfer between both components and gravitation forces from 
companions. Recently, Liakos \& Niarchos (2015, 2016a) showed that there is a threshold in the orbital period of $\sim$ 13 days 
below which the pulsations are influenced by the binarity. In eccentric-orbit binaries, some pulsations can be excited by 
tidal interaction. The signature of the tidally excited modes is the frequencies at multiples of the orbital frequency 
(Welsh et al. 2011; Hambleton et al. 2013). 

This paper is the fifth contribution in a series of detailed studies for pulsating stars in the {\it Kepler} EBs 
(Lee et al. 2014, 2016a,b; Lee 2016). We present that KIC 11401845 (R.A.$_{2000}$ = 19$^{\rm h}$25$^{\rm m}$11$\fs275$; 
decl.$_{2000}$ = +49$^{\circ}$14${\rm '}$40$\farcs$09; $K_{\rm p}$ = $+$14.355; $g$ = $+$14.443; $g-r$ = $+$0.118) is 
a detached EB exhibiting multiperiodic pulsations and light-travel-time (LTT) delay. The system was announced to be an EB 
pulsating at frequencies of 13$-$25 d$^{-1}$ by Gaulme \& Guzik (2014).

\section{{\it KEPLER} PHOTOMETRY AND LIGHT-CURVE SYNTHESIS}

The {\it Kepler} data of KIC 11401845 were obtained in a long cadence mode of 29.42 minutes during Quarters 10 and 12. 
We used the SAP (Simple Aperture Photometry) time-series data in Data Release 25 retrieved from the {\it Kepler} Data 
Archive\footnote{http://archive.stsci.edu/kepler/}. The raw data were detrended by using second-order polynomials that were 
separately applied to each quarter. As eclipses influence the detrending process, we fitted the polynomial to 
only the outside-eclipse part of the light curve (e.g., Hambleton et al. 2013). The flux measurements were converted to 
a magnitude scale by requiring a {\it Kepler} magnitude of +14.355 at maximum light. The detrended {\it Kepler} data are 
displayed in Figure 1. The shape of the light curve indicates a significant temperature difference between the binary components 
and an ellipsoidal variation due to tidal distortions.

To derive the binary parameters of this system, all {\it Kepler} data were analyzed using the Wilson-Devinney binary program 
(Wilson \& Devinney 1971, van Hamme \& Wilson 2007; hereafter W-D). This synthesis was performed in a way similar to that 
for the pulsating EBs V404 Lyr (Lee et al. 2014) and KIC 6220497 (Lee et al. 2016a). The effective temperature 
of the hotter and more massive star was set to be 7590 K from the {\it Kepler} Input Catalogue (KIC; Kepler Mission Team 2009). 
The logarithmic bolometric ($X_{1,2}$) and monochromatic ($x_{1,2}$) limb-darkening coefficients were interpolated from 
the values of van Hamme (1993). In Table 1, the parentheses signify the adjusted parameters. In this article, the subscripts 
1 and 2 refer to the primary and secondary components being eclipsed at orbital phases 0.0 (Min I) and 0.5 (Min II), respectively. 

There has been neither a light-curve solution nor spectroscopic mass ratio ($q$) for KIC 11401845 so far. Thus, we conducted 
a photometric $q$-search procedure that calculates a series of models with varying $q$ from 0 to 1. For each assumed 
mass ratio, the W-D code was applied for various modes but converged satisfactorily only when detached mode 2 were used. 
The weighted sum of the squared residuals, $\sum W(O-C)^2$, reached a minimum around $q$ = 0.07, which was adopted as 
the initial value and thereafter adjusted to derive the photometric solutions. The result is given as Model 1 in the second 
and third columns of Table 1. The synthetic light curve appears as the blue solid curve in Figure 1, and the corresponding 
light residuals are plotted as the gray `x' symbols in the figure. In all the procedures, we considered an orbital eccentricity 
($e$) and a third light ($l_3$) as additional free parameters. Both searches led to values for the two parameters which were 
zero within their errors, which implies that KIC 11401845 has negligible eccentricity. We obtained the errors for the adjustable 
parameters by splitting the {\it Kepler} data into 79 segments at intervals of an orbital period and analyzing them separately 
(Koo et al. 2014). In Table 1, the error of each parameter is its standard deviation computed from this process. 

Absolute parameters for KIC 11401845 can be computed from our light-curve solutions in Table 1 and from the correlations 
between spectral type (temperature) and stellar mass. The surface temperature ($T_1$ = 7590 K) of the primary star corresponds 
to a normal dwarf one with a spectral type of $\sim$A8V. Assuming that each component has a temperature error of 200 K, 
the primary's mass was estimated to be $M_1$ = 1.70$\pm$0.08 $M_\odot$ from Harmanec's (1988) empirical relation. We calculated 
the absolute dimensions for each component given in the last part of Table 1. Here, the luminosity ($L$) and bolometric 
magnitudes ($M_{\rm bol}$) were derived by using $T_{\rm eff}$$_\odot$ = 5780 K and $M_{\rm bol}$$_\odot$ = +4.73. 
The bolometric corrections (BCs) were obtained from the expression between $\log T_{\rm eff}$ and BC given by Torres (2010). 

Considering the temperature error of the primary star, we carried out the light-curve synthesis for 7390 K and 7790 K. 
The binary parameters from the two limits are in satisfactory agreement with those from $T_1$ = 7590 K, except for 
the secondary's temperature ($T_2$). In the three models, the temperature ratios ($T_2$/$T_1$) of both components 
are consistent with each other within their errors. We can see that the adopted $T_1$ does not affect the results presented 
in this paper.

\section{LIGHT RESIDUALS AND PULSATIONAL CHARACTERISTICS}

From the temperature ($T_1$) and surface gravity ($\log$ $g_1$) given in Table 1, the primary star of KIC 11401845 resides 
within the $\delta$ Sct instability strip and, hence, it would be a candidate for $\delta$ Sct pulsations. 
For more reliable frequency analysis, we followed the procedure described by Lee (2016). First, we divided the observed 
{\it Kepler} data of KIC 11401845 into 79 subsets as before and modeled each light curve with the W-D code by adjusting 
only the ephemeris epoch ($T_0$) in Model 1 of Table 1. Second, the corresponding residuals from the whole datasets were 
applied to multiple frequency analyses in the range from 0 to the Nyquist limit of $f_{\rm Ny}$ = 24.4 day$^{-1}$ using 
the PERIOD04 program (Lenz \& Breger 2005). Because the binary components block each other's lights during eclipses, we used only 
the outside-eclipse residuals (orbital phases 0.07$-$0.43 and 0.57$-$0.93). According to the successive prewhitening procedures, 
we detected the frequencies with signal to noise amplitude (S/N) ratios larger than 4.0 (Breger et al. 1993). Third, we solved 
the pulsation-subtracted data after removing the pulsations from the observed data. As a result, new binary parameters were 
obtained, and they were used to reanalyze the 79 light curves in the first stage.

This process was repeated three times until the detected frequencies were unchanged. Final binary parameters are given as 
Model 2 in the fourth and fifth columns of Table 1, and the pulsation-subtracted data and model light curve are plotted in 
Figure 2. We can see that the physical parameters for Model 2 are consistent with those for Model 1. Figure 3 shows 
the light residuals after removal of the binary effects from the observed {\it Kepler} data. We detected a total of 
23 frequencies larger than the empirical threshold of S/N = 4.0. The amplitude spectra before and after prewhitening 
the first 10 frequencies and then all 23 frequencies are shown in the first to third panels of Figure 4, respectively. 
The detailed result from the frequency analysis is listed in Table 2, where the frequencies are given in order of detection 
and the noises are calculated in the range of 5 day$^{-1}$ around each frequency. The uncertainties in the table were obtained 
according to Kallinger et al. (2008). The synthetic curve computed from the 23-frequency fit is displayed in the lower panel 
of Figure 3. 

As listed in Table 2, four in the low-frequency region (0.4$-$3.8 day$^{-1}$) and 19 in the high-frequency region 
(13.7$-$23.8 day$^{-1}$) were derived from the multiple frequency analyses of the outside-eclipse light residuals. 
Within the frequency resolution of 0.00545 day$^{-1}$ (Loumos \& Deeming 1978), we searched the frequencies for possible 
harmonic and combination terms. The result is given in the last column of Table 2. We think that the $f_{11}$ to $f_{23}$ 
frequencies mainly arise from combination frequencies, some of which can be caused by imperfect removal of the binary effects 
in the observed data. On the other hand, the high-frequency signals close to the Nyquist limit can be reflections of 
real frequencies (2$f_{\rm Ny}-f_i$) higher than $f_{\rm Ny}$ (Murphy et al. 2013; Lee et al. 2016b). High-cadence photometry 
is needed to separate the Nyquist aliases from the detected frequencies of KIC 11401845.

\section{ECLIPSE TIMING VARIATION AND ITS IMPLICATION}

For an orbital period study of KIC 11401845, we determined 150 minimum times and their uncertainties from the observations with 
the method of Kwee \& van Woerden (1956). These minima are listed in columns (1)--(5) of Table 3, where we present the cycle 
numbers and $O$--$C_1$ residuals calculated with the light elements ($T_0$ and $P$) for Model 2 in Table 1. The resultant eclipse 
timing diagram is displayed at the top panel of Figure 5. As shown in the figure, the timing residuals from the primary 
(filled circle) and secondary (open circle) eclipses do not agree with each other, which could be caused by the light variations 
due to the multiperiodic pulsations of the primary star. Thus, we recalculated the minimum times from the eclipse light curve 
after subtracting the 23 frequencies detected in Section 3 from the observed {\it Kepler} data. The results are given in columns 
(6)--(8) of Table 3 and are illustrated in the middle panel of Figure 5. 

As displayed in Figure 5, the large discrepancy between Min I and Min II is shown more clearly in the pulsation-subtracted data.
This discrepancy can result from the time difference between the primary and secondary eclipses due to LTT in a binary with 
unequal masses (Barlow et al. 2012; Parsons et al. 2014). The R{\o}mer delay is given by Kaplan (2010), as follows:
\begin{equation}
 \Delta t_{\rm LTT} = {{P K_2} \over {\pi c}} (1-q),
\end{equation}
where $P$ is the orbital period, $K_2$ is the radial velocity (RV) semi-amplitude of the secondary star, and c is the speed of 
light. From the Model 2 parameters in Table 1, we derived the velocities ($K_1$ = 13 km s$^{-1}$ and $K_2$ = 187 km s$^{-1}$) 
of the binary components (Hilditch 2001). The time delay of $\Delta t_{\rm LTT}$ = 34$\pm$1 s was obtained by applying the values 
of $K_2$, $P$, and $q$ to equation (1). 

In order to examine this possibility, we computed the secondary eclipses related to one half period after the primary eclipses 
in the pulsation-subtracted data and then plotted the difference ($\Delta t_{\rm SE}$) between the measured and computed 
secondary times in the bottom panel of Figure 5. As shown in the figure, the mean value is offset from zero and gives 
a time delay of $\Delta t_{\rm SE}$ = 56$\pm$17 s in the secondary eclipse. This value is calculated to be $\Delta t_{\rm SE}$ 
= 37$\pm$27 s in the observed {\it Kepler} data including pulsations. Within their errors, the time delays of $\Delta t_{\rm SE}$ 
are in satisfactory accord with the predicted delay of $\Delta t_{\rm LTT}$. On the other hand, if KIC 11401845 is in an eccentric orbit, 
$\Delta t_{\rm SE}$ might be affected by the time shift of $\Delta t_{\rm e}$ in the secondary eclipse due to a non-zero eccentricity:
\begin{eqnarray}
 \Delta t_{\rm e} \simeq {{2Pe} \over {\pi}} \cos {\omega}, \\
 \Delta t_{\rm SE} \simeq \Delta t_{\rm e} + \Delta t_{\rm LTT}, 
\end{eqnarray}
where $e$ and $\omega$ are the eccentricity and the argument of periastron, respectively. Using the equations (2) and (3), 
$e \cos {\omega} \simeq$ 0.00003 for the observed data and $e \cos {\omega} \simeq$ 0.00019 for the pulsation-subtracted data.

\section{DISCUSSION AND CONCLUSIONS}

We have studied the physical properties of KIC 11401845, based on the {\it Kepler} data made during Quarters 10 and 12. 
The light curve of this system displays multiperiodic pulsations, superimposed on binary effects. To examine whether the binary 
parameters are affected by the pulsations, we analyzed individually the observed and pulsation-subtracted {\it Kepler} data 
with the W-D code. As listed in Table 1, the photometric solutions for the two datasets are in good agreement with each other, 
which means that the pulsations cause little impact on the light-curve parameters. Our light-curve synthesis shows that 
KIC 11401845 is a short-period detached EB with a very small mass ratio of about 0.07. The primary and secondary components 
fill $F_1$ = 45 \% and $F_2$ = 99 \% of their limiting lobe, respectively, where the filling factor 
$F_{1,2} = \Omega_{\rm in} / \Omega_{1,2}$. With its small $q$ and short $P$, our program target closely resembles 
the two {\it Kepler} pulsating EBs KIC 10661783 (Lehmann et al. 2013) and KIC 8262223 (Guo et al. 2016), which are detached 
binaries with characteristics of the R CMa-type stars (Budding \& Butland 2011; Lee et al. 2016b) among Algols. A comparison of 
the KIC 11401845 parameters with the mass-radius, mass-luminosity, and Hertzsprung-Russell (HR) diagrams (Ibano\v{g}lu et al. 2006) 
shows that the primary component resides within the main-sequence band. On the contrary, the low mass secondary is highly evolved 
and its radius and luminosity are remarkably oversized and overluminous compared to main-sequence stars of the same mass. These 
suggest that the initial more massive star becomes the present secondary by losing most of its own mass via mass transfer to 
the companion (present primary) and stellar wind (Hong et al. 2015; Guo et al. 2016).

In order to detect the pulsation frequencies of KIC 11401845, multiple frequency analyses were applied to the out-of-eclipse 
residuals after removing the binarity effects from the observed {\it Kepler} data. We found 23 frequencies with S/N ratios 
larger than 4.0 in two regions: 0.4$-$3.8 day$^{-1}$ and 13.7$-$23.8 day$^{-1}$. Among these, four ($f_4$, $f_6$, $f_9$, $f_{15}$) 
in the low frequency region are frequencies at exact multiples of the orbital frequency, $f_{\rm orb}$ = 0.46267 day$^{-1}$. 
The orbital harmonics can be attributed to tidally excited modes, which occur when the orbital frequency is close to 
a stellar eigenfrequency in a binary star (Welsh et al. 2011; Hambleton et al. 2013). On the contrary, the other frequencies 
are strongly reminiscent of the $p$-mode pulsations known in EBs (e.g., Southworth et al. 2011; Lee et al. 2016b). The ratios of 
the pulsational to orbital periods in the high-frequency region were calculated to be $P_{\rm pul}/P_{\rm orb}$ = 0.020$-$0.034, 
which is within the upper limit of 0.09 for $\delta$ Sct stars in binaries (Zhang et al. 2013). We calculated the pulsation 
constants by applying the Model 2 parameters in Table 1 to the equation of 
$\log Q_i = -\log f_i + 0.5 \log g + 0.1M_{\rm bol} + \log T_{\rm eff} - 6.456$ (Petersen \& J\o rgensen 1972). The result is 
listed in the sixth column of Table 2. The $Q$ values of 0.018$-$0.031 d correspond to $p$ modes of $\delta$ Sct type. The period 
ratios, the pulsation constants, and the position on the HR diagram reveal that the primary component is a $\delta$ Sct variable.
The $\delta$ Sct pulsations of KIC 11401845 match well the correlations between the pulsation periods and other parameters 
(binary periods, surface gravities of pulsators, and gravitational forces from companions) updated by Liakos \& Niarchos (2016b).

We measured the minimum epochs of the primary and secondary eclipses from the observed and pulsation-subtracted data. 
Inspecting them in detail, we found delays of about 37 s and 56 s in the arrival times of the secondary eclipses relative to 
the primary eclipses in the same order. The values are consistent with the expected time delay of $\Delta t_{\rm LTT}$ = 34 s 
across the binary orbit. This indicates that the LTT delay is the main cause of the time discrepancy between both eclipses, 
which is the first detection of this effect in EBs consisting of non-compact components. One might imagine that the time delay 
of the secondary eclipse could be apportioned between the LTT delay and a non-zero eccentricity. The result presented in 
this paper limits the eccentricity of KIC 11401845 to $e < 0.0002$. When the high-resolution spectra are made, they will help to 
determine the RV semi-amplitudes ($K_{1,2}$) and mass ratio ($q$) of the binary star and hence to derive its small eccentricity 
($e$). Because the system is a faint pulsating EB with a short orbital period, 8$-$10 m class telescopes are required to 
measure its accurate double-lined RVs.

\acknowledgments{ }
This paper includes data collected by the {\it Kepler} mission. {\it Kepler} was selected as the 10th mission of the Discovery Program. 
Funding for the {\it Kepler} mission is provided by the NASA Science Mission directorate. We have used the Simbad database maintained 
at CDS, Strasbourg, France. This work was supported by the KASI grant 2016-1-832-01. Work by K. Hong was supported by Basic Science 
Research Program through the National Research Foundation of Korea (NRF) funded by the Ministry of Education (grant number: NRF-2016R1A6A3A01007139).

\clearpage
\begin{figure}
\includegraphics[scale=0.85]{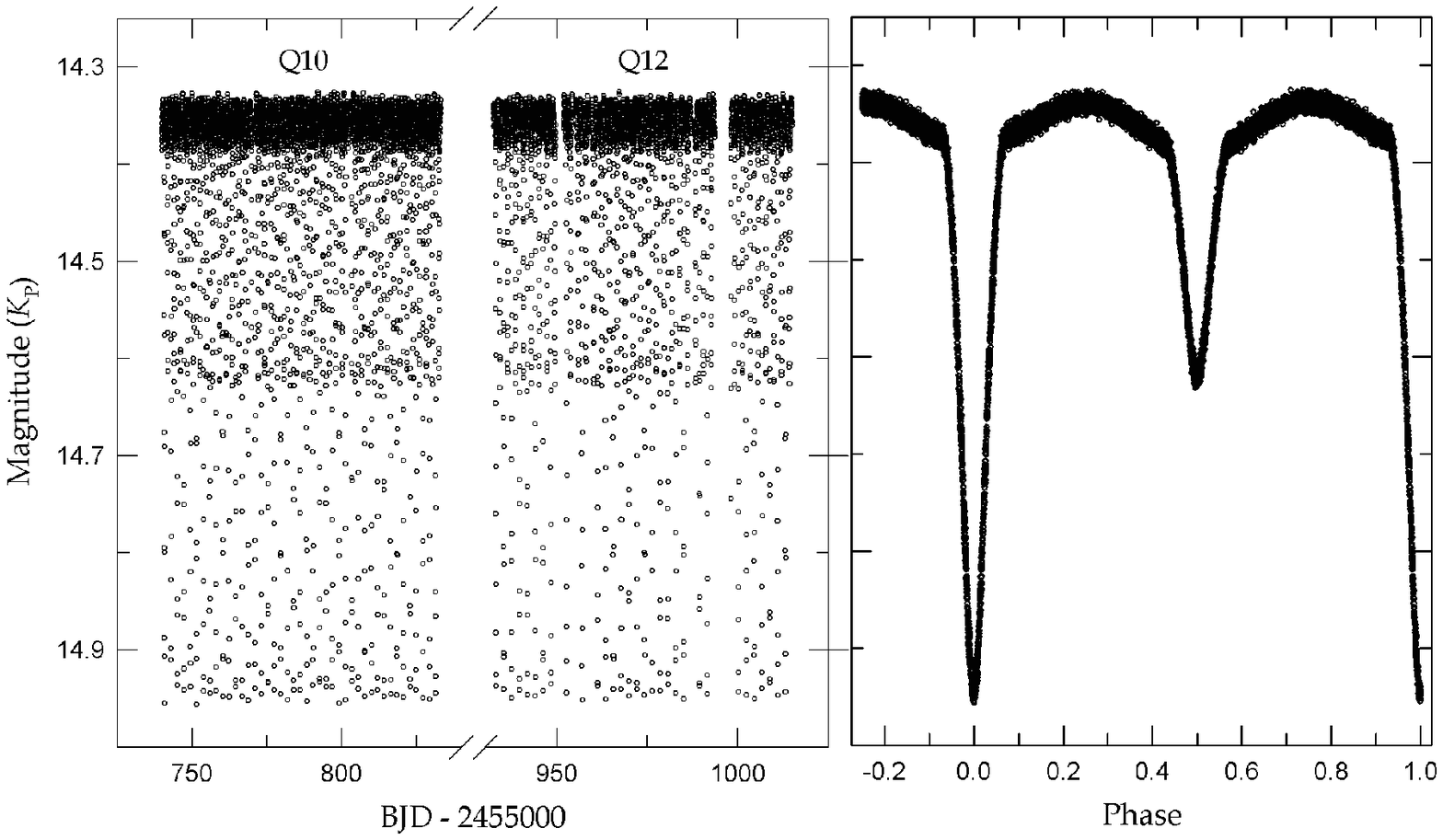}
\caption{Detrended {\it Kepler} data of KIC 11401845 distributed in BJD (left panel) and orbital phase (right panel). }
\label{Fig1}
\end{figure}

\begin{figure}
\includegraphics[]{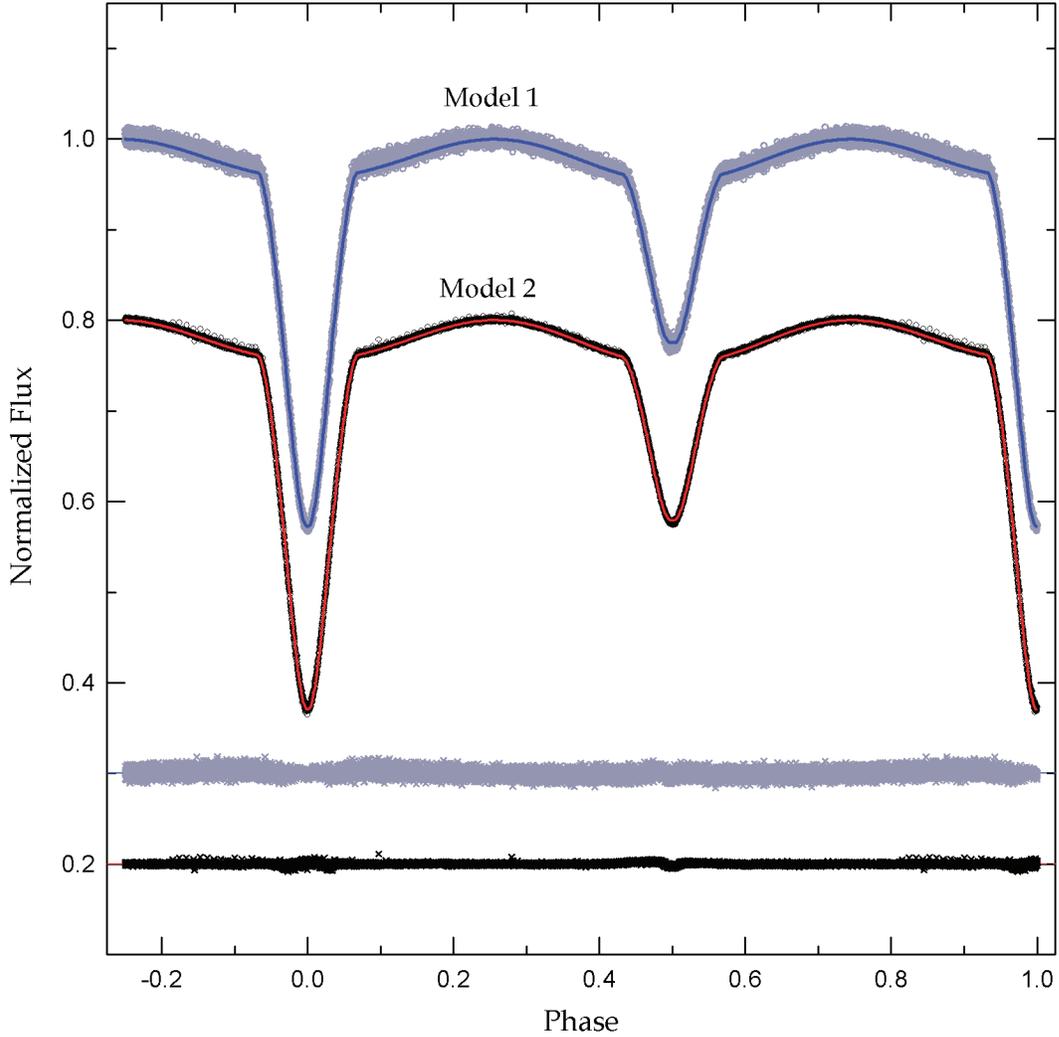}
\caption{Binary light curve before (gray circle) and after (black circle) subtracting the pulsation signatures from 
the {\it Kepler} data. The blue and red solid curves are computed with the Model 1 and Model 2 parameters of Table 1, 
respectively. The corresponding residuals from the fits are offset from zero and plotted at the bottom in the same order 
as the light curves. In the light residuals from Model 2, some feature is certainly visible during the times of 
the secondary eclipse, which may come from insufficient removal of the pulsation effects in the orbital phases. }
\label{Fig2}
\end{figure}

\begin{figure}
\includegraphics[]{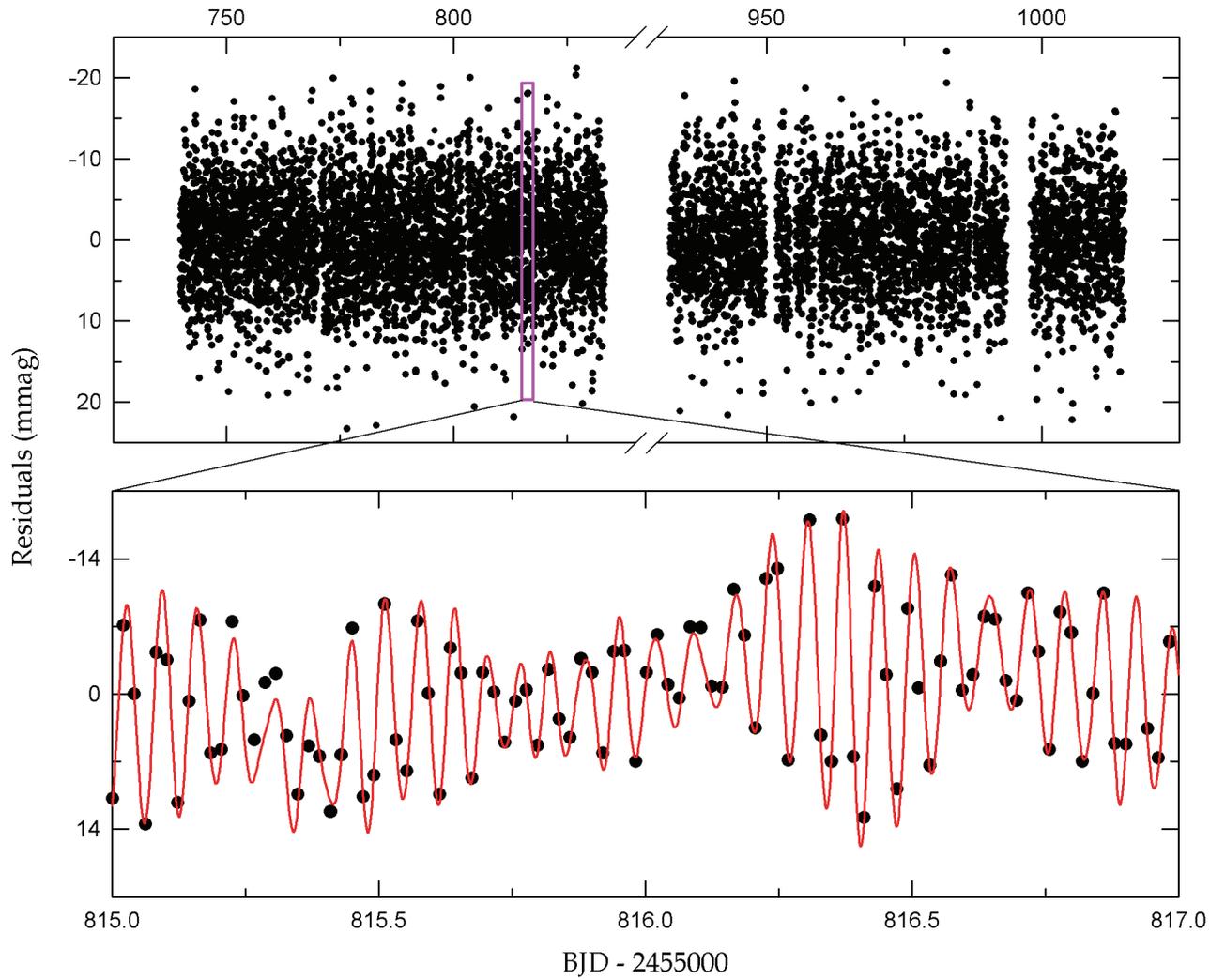}
\caption{Light residuals after removing the binarity effects from the {\it Kepler} light curve. The lower panel presents 
a short section of the residuals marked by the inset box in the upper panel. The synthetic curve is computed from 
the 23-frequency fit to the data. }
\label{Fig3}
\end{figure}

\begin{figure}
\includegraphics[]{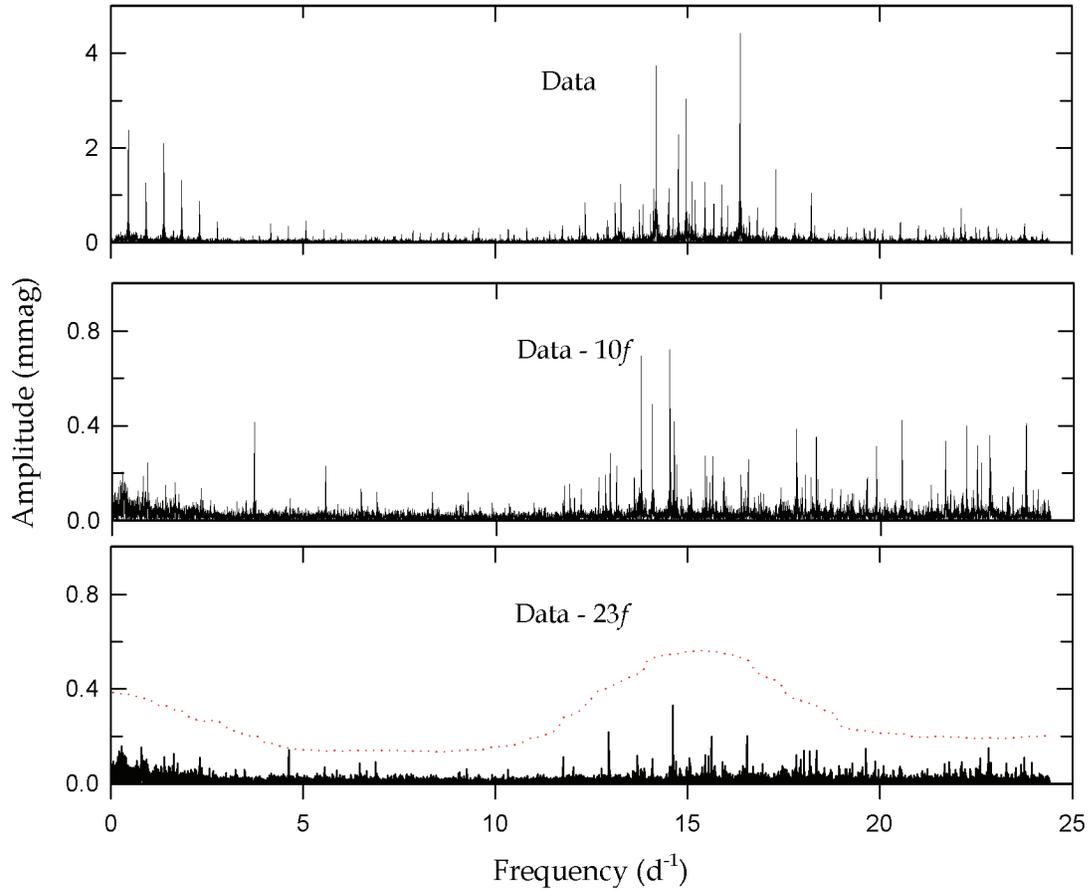}
\caption{Amplitude spectra before (top panel) and after prewhitening the first 10 frequencies (middle panel) and 
all 23 frequencies (bottom panel) from the PERIOD04 program for the entire outside-eclipse residual data. 
The dotted line at the bottom panel corresponds to four times the noise spectrum, which was calculated for each frequency 
in an equidistant step of 0.1 day$^{-1}$.}
\label{Fig4}
\end{figure}

\begin{figure}
\includegraphics[]{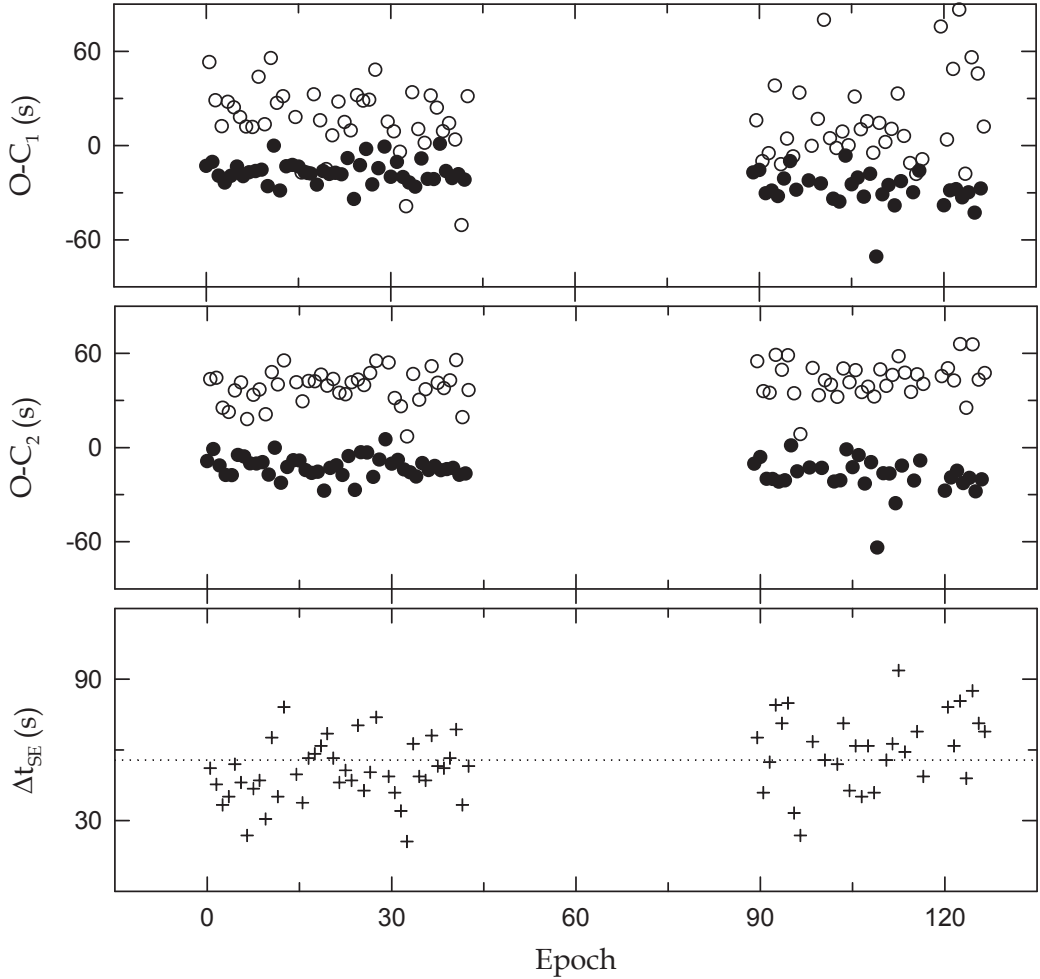}
\caption{$O$--$C$ diagrams of the minimum times measured from the observed (top panel) and pulsation-subtracted (middle panel) 
{\it Kepler} data. The filled and open circles represent the primary and secondary minima, respectively. Bottom panel shows 
the time delay ($\Delta t_{\rm SE}$) of the secondary eclipses related to one half period after the primary eclipses in 
the pulsation-subtracted data. The dotted line refers to the mean value of $\Delta t_{\rm SE}$ = 56$\pm$17 s. }
\label{Fig5}
\end{figure}

\clearpage
\begin{deluxetable}{lcccccccc}
\tablewidth{0pt} 
\tablecaption{Binary Parameters of KIC 11401845}
\tablehead{
\colhead{Parameter}                      & \multicolumn{2}{c}{Model 1$\rm ^a$}         && \multicolumn{2}{c}{Model 2$\rm ^b$}         \\ [1.0mm] \cline{2-3} \cline{5-6} \\[-2.0ex]
                                         & \colhead{Primary} & \colhead{Secondary}     && \colhead{Primary} & \colhead{Secondary}                                                  
}                                                                                                                                     
\startdata                                                                                                                            
$T_0$ (BJD)                              & \multicolumn{2}{c}{2,455,740.80720(13)}     && \multicolumn{2}{c}{2,455,740.80729(5)}      \\
$P$ (day)                                & \multicolumn{2}{c}{2.1613895(18)}           && \multicolumn{2}{c}{2.1613910(6)}            \\
$q$                                      & \multicolumn{2}{c}{0.0699(16)}              && \multicolumn{2}{c}{0.0695(10)}              \\
$i$ (deg)                                & \multicolumn{2}{c}{85.45(21)}               && \multicolumn{2}{c}{85.10(12)}               \\
$T$ (K)                                  & 7590              & 6217(26)                && 7590              & 6154(19)                \\
$\Omega$                                 & 4.150(29)         & 1.879(7)                && 4.138(21)         & 1.873(4)                \\
$\Omega_{\rm in}$                        & \multicolumn{2}{c}{1.861}                   && \multicolumn{2}{c}{1.860}                   \\
$A$                                      & 1.0               & 0.5                     && 1.0               & 0.5                     \\
$g$                                      & 1.0               & 0.32                    && 1.0               & 0.32                    \\
$X$, $Y$                                 & 0.671, 0.199      & 0.632, 0.232            && 0.671, 0.199      & 0.633, 0.229            \\
$x$, $y$                                 & 0.599, 0.238      & 0.629, 0.280            && 0.599, 0.238      & 0.633, 0.278            \\
$L$/($L_{1}$+$L_{2}$)                    & 0.7905(12)        & 0.2095                  && 0.7940(8)         & 0.2060                  \\
$r$ (pole)                               & 0.2450(13)        & 0.1638(29)              && 0.2457(10)        & 0.1651(19)              \\
$r$ (point)                              & 0.2474(14)        & 0.2083(78)              && 0.2481(10)        & 0.2136(56)              \\
$r$ (side)                               & 0.2470(14)        & 0.1695(32)              && 0.2476(10)        & 0.1710(21)              \\
$r$ (back)                               & 0.2473(14)        & 0.1928(49)              && 0.2480(10)        & 0.1956(32)              \\
$r$ (volume)$\rm ^c$                     & 0.2464(14)        & 0.1753(43)              && 0.2471(10)        & 0.1771(29)              \\ 
$\sum W(O-C)^2$                          & \multicolumn{2}{c}{0.0050}                  && \multicolumn{2}{c}{0.0016}                  \\ [1.0mm]
\multicolumn{6}{l}{Absolute parameters:}                                                                                              \\            
$M$ ($M_\odot$)                          & 1.70(8)           & 0.119(7)                && 1.70(8)           & 0.118(6)                \\
$R$ ($R_\odot$)                          & 2.11(5)           & 1.50(5)                 && 2.12(4)           & 1.52(4)                 \\
$\log$ $g$ (cgs)                         & 4.02(3)           & 3.16(4)                 && 4.02(3)           & 3.15(3)                 \\
$L$ ($L_\odot$)                          & 13(2)             & 3.0(4)                  && 13(2)             & 3.0(4)                  \\
$M_{\rm bol}$ (mag)                      & 1.92(12)          & 3.53(16)                && 1.92(12)          & 3.55(15)                \\
BC (mag)                                 & 0.03(1)           & $-$0.02(2)              && 0.03(1)           & $-$0.03(2)              \\
$M_{\rm V}$ (mag)                        & 1.89(12)          & 3.55(16)                && 1.89(12)          & 3.58(15)                \\
\enddata
\tablenotetext{a}{Result from the observed data.}
\tablenotetext{b}{Result from the pulsation-subtracted data.}
\tablenotetext{c}{Mean volume radius.}
\end{deluxetable}

\begin{deluxetable}{lrccccc}
\tabletypesize{\small}
\tablewidth{0pt}
\tablecaption{Multiple Frequency Analysis of KIC 11401845}
\tablehead{
             & \colhead{Frequency}    & \colhead{Amplitude} & \colhead{Phase} & \colhead{S/N}  & \colhead{$Q$}     & \colhead{Remark}                \\
             & \colhead{(day$^{-1}$)} & \colhead{(mmag)}    & \colhead{(rad)} &                & \colhead{(days)}  &
}                                                                                                                                
\startdata                                                                                                                       
$f_{1}$      & 16.37811$\pm$0.00004   & 4.41$\pm$0.23       & 5.74$\pm$0.15   & 33.13          & 0.026             &                                 \\
$f_{2}$      & 14.19519$\pm$0.00004   & 3.74$\pm$0.23       & 2.06$\pm$0.18   & 27.93          & 0.030             &                                 \\
$f_{3}$      & 14.97171$\pm$0.00005   & 3.18$\pm$0.24       & 5.18$\pm$0.22   & 22.85          & 0.028             &                                 \\
$f_{4}$      &  0.46268$\pm$0.00003   & 3.54$\pm$0.16       & 5.73$\pm$0.13   & 37.42          &                   & $f_{\rm orb}$                   \\
$f_{5}$      & 14.76672$\pm$0.00007   & 2.28$\pm$0.24       & 2.63$\pm$0.30   & 16.56          & 0.029             &                                 \\
$f_{6}$      &  1.85072$\pm$0.00005   & 1.81$\pm$0.13       & 5.33$\pm$0.21   & 23.56          &                   & $4f_{\rm orb}$                  \\
$f_{7}$      & 14.12286$\pm$0.00014   & 1.17$\pm$0.23       & 3.79$\pm$0.57   &  8.78          & 0.030             &                                 \\
$f_{8}$      & 15.20396$\pm$0.00019   & 0.90$\pm$0.24       & 4.42$\pm$0.78   &  6.45          & 0.028             &                                 \\
$f_{9}$      &  2.77609$\pm$0.00007   & 1.11$\pm$0.11       & 0.95$\pm$0.30   & 16.92          &                   & $6f_{\rm orb}$                  \\
$f_{10}$     & 22.12255$\pm$0.00008   & 0.73$\pm$0.08       & 1.13$\pm$0.33   & 14.99          & 0.019             &                                 \\
$f_{11}$     & 14.50213$\pm$0.00023   & 0.74$\pm$0.23       & 3.76$\pm$0.93   &  5.41          & 0.029             & $3f_3-2f_8$                     \\
$f_{12}$     & 13.75268$\pm$0.00021   & 0.69$\pm$0.20       & 3.11$\pm$0.85   &  5.90          & 0.031             &                                 \\
$f_{13}$     & 14.04653$\pm$0.00030   & 0.55$\pm$0.23       & 1.53$\pm$1.20   &  4.16          & 0.030             & $f_3-2f_{\rm orb}$              \\
$f_{14}$     & 20.54260$\pm$0.00016   & 0.41$\pm$0.09       & 5.00$\pm$0.64   &  7.87          & 0.021             & $f_1+9f_{\rm orb}$              \\
$f_{15}$     &  3.70145$\pm$0.00012   & 0.52$\pm$0.09       & 2.88$\pm$0.48   & 10.35          &                   & $8f_{\rm orb}$                  \\
$f_{16}$     & 22.22705$\pm$0.00015   & 0.40$\pm$0.08       & 3.22$\pm$0.61   &  8.18          & 0.019             & $2f_7-13f_{\rm orb}$            \\
$f_{17}$     & 23.77410$\pm$0.00015   & 0.40$\pm$0.08       & 0.08$\pm$0.62   &  8.11          & 0.018             &                                 \\
$f_{18}$     & 17.79523$\pm$0.00029   & 0.38$\pm$0.15       & 4.34$\pm$1.19   &  4.21          & 0.024             &                                 \\
$f_{19}$     & 22.82621$\pm$0.00017   & 0.36$\pm$0.08       & 3.22$\pm$0.68   &  7.34          & 0.019             &                                 \\
$f_{20}$     & 21.67877$\pm$0.00018   & 0.35$\pm$0.08       & 2.34$\pm$0.71   &  7.03          & 0.020             &                                 \\
$f_{21}$     & 18.31316$\pm$0.00030   & 0.33$\pm$0.14       & 4.19$\pm$1.23   &  4.07          & 0.023             &                                 \\
$f_{22}$     & 22.50255$\pm$0.00019   & 0.31$\pm$0.08       & 6.20$\pm$0.77   &  6.49          & 0.019             &                                 \\
$f_{23}$     & 19.87857$\pm$0.00021   & 0.31$\pm$0.09       & 3.96$\pm$0.87   &  5.79          & 0.021             &                                 \\
\enddata
\end{deluxetable}

\begin{deluxetable}{lcccccclcc}
\tabletypesize{\small}  
\tablewidth{0pt}
\tablecaption{Eclipse Timings Measured from Both Datasets Including and Removing the pulsations }
\tablehead{
\multicolumn{3}{c}{Observed Data}                         &&                 &                && \multicolumn{3}{c}{Pulsation-Subtracted Data}               \\ [1.0mm] \cline{1-3} \cline{8-10} \\[-2.0ex]
\colhead{BJD}    & \colhead{Error} & \colhead{$O$--$C_1$} && \colhead{Epoch} & \colhead{Min}  && \colhead{BJD}    & \colhead{Error} & \colhead{$O$--$C_2$} 
}                                                                         
\startdata                                                                
2,455,740.80714  & $\pm$0.00031    & $-$0.000150          &&      0.0        & I              && 2,455,740.80719  & $\pm$0.00014    & $-$0.000100            \\
2,455,741.88860  & $\pm$0.00120    & $+$0.000615          &&      0.5        & II             && 2,455,741.88849  & $\pm$0.00028    & $+$0.000504            \\
2,455,742.96856  & $\pm$0.00064    & $-$0.000121          &&      1.0        & I              && 2,455,742.96867  & $\pm$0.00039    & $-$0.000011            \\
2,455,744.04971  & $\pm$0.00083    & $+$0.000334          &&      1.5        & II             && 2,455,744.04989  & $\pm$0.00031    & $+$0.000514            \\
2,455,745.12985  & $\pm$0.00038    & $-$0.000222          &&      2.0        & I              && 2,455,745.12994  & $\pm$0.00025    & $-$0.000132            \\
2,455,746.21091  & $\pm$0.00064    & $+$0.000142          &&      2.5        & II             && 2,455,746.21106  & $\pm$0.00020    & $+$0.000293            \\
2,455,747.29119  & $\pm$0.00050    & $-$0.000273          &&      3.0        & I              && 2,455,747.29126  & $\pm$0.00020    & $-$0.000203            \\
2,455,748.37248  & $\pm$0.00162    & $+$0.000322          &&      3.5        & II             && 2,455,748.37242  & $\pm$0.00034    & $+$0.000262            \\
2,455,749.45263  & $\pm$0.00030    & $-$0.000224          &&      4.0        & I              && 2,455,749.45265  & $\pm$0.00028    & $-$0.000204            \\
2,455,750.53383  & $\pm$0.00052    & $+$0.000281          &&      4.5        & II             && 2,455,750.53397  & $\pm$0.00019    & $+$0.000421            \\
\enddata
\tablecomments{This table is available in its entirety in machine-readable and Virtual Observatory (VO) forms in the online journal. 
A portion is shown here for guidance regarding its form and content.}
\end{deluxetable}

\end{document}